\documentclass[prb,showpacs,twocolumn,floatfix]{revtex4-1}
\usepackage{graphicx}
\usepackage{dcolumn}
\usepackage{bm}
\usepackage{amssymb,amsmath}
\usepackage{color}

\begin{document}
\title{Critical stress statistics and a fold catastrophe in intermittent crystal plasticity}
\author{P. M. Derlet}
\email{Peter.Derlet@psi.ch} 
\affiliation{Condensed Matter Theory Group, Paul Scherrer Institute, CH-5232 Villigen PSI, Switzerland}
\author{R. Maa{\ss}}
\affiliation{Department of Materials Science and Engineering, University of Illinois at Urbana Champaign, 1304 West Green Street, Urbana, Illinois 61801, USA }
\date{\today}

\begin{abstract}
The statistics and origin of the first discrete plastic event in a one dislocation dynamics simulation are studied. This is done via a linear stability analysis of the evolving dislocation configuration up to the onset of irreversible plasticity. It is found, via a fold catastrophe, the dislocation configuration prior to loading directly determines the stress at which the plastic event occurs and that between one and two trigger dislocations are involved. The resulting irreversible plastic strain arising from the instability is found to be highly correlated with these triggering dislocations.
\end{abstract}
\maketitle
\section{Introduction}

The transition from elastic to macroscopic plastic deformation in metals is a discrete process mediated by intermittent dislocation activity. Historically, such discrete plastic activity was seen in bulk metallic systems by Beeker and Orowan~\cite{Becker1932} and later by Tinder and co-workers~\cite{Tinder1964,Tinder1973} who used a torsion device, with ultra-high strain resolution, to show that plasticity well below yield was characterized by discrete plastic events separated by regions of near perfect elasticity. Modern developments in the instrumentation of a deformation experiment have allowed this phenomenon to be studied in more detail. For example, intermittent plasticity is routinely seen via in nano-identation as a ``pop-in'' event~\cite{Shim2008,Morris2011} and in the flat-punch indention of a focus-ion-beam milled micron-sized crystal, as first introduced by Uchic and co-workers~\cite{Uchic2004}.

One interesting aspect of the above phenomena is that as the size of the plastically deforming region reduces the stress scale at which the discrete plasticity occurs increases. Such an effect is best known through the paradigm of ``smaller-is-stronger'', first coined in the micron-deformation experiments of ref.~\cite{Uchic2004} (see ref.~\cite{Greer2011} and references therein). A size effect in the pop-in stress as a function of spherical indenter size has also been seen in refs.~\cite{Shim2008,Morris2011}. For the case of the micron-deformation experiments, the understanding of the smaller-is-stronger effect has primarily focused on new dislocation based mechanisms becoming operative as the plastically deforming volume reduces --- mechanisms that depend explicitly on the presence of open boundary conditions --- see ref.~\cite{Elawady2015} and references therein. Despite the possibility of a change-of-mechanism approach, it has long been recognised that with decreasing sample size, the overall evolution of deformation becomes increasingly stochastic with particular realizations of nominally similar deforming volumes exhibiting a range of responses. Indeed, this aspect has been used to explain the size effect in nano-indentation~\cite{Morris2011}, where only for the very smallest indenter sizes a change in mechanism was needed~\cite{Phani2013}.

Under the assumption that the general size-effect still involves bulk plasticity mechanisms, the present authors have employed an extreme-value statistics approach to rationalize a size effect in the stress scale at which discrete plasticity occurs~\cite{Derlet2015a,Maass2015a,Derlet2015b}. More recently, this framework has been applied to study the statistics of the critical stress at which the very first plastic event occurs~\cite{Derlet2016}. This was done for the experimental nano-indentation data of Morris {\em et al}~\cite{Morris2011} and a simplified dislocation dynamics simulation under periodic boundary conditions~\cite{Derlet2013}.

In ref.~\cite{Derlet2016}, the statistics of the stress at which the first plastic event occurs is described by the extreme value statistics of a master distribution of critical stresses, $P[\sigma]$, which characterizes the initial plastic response of the material. Within this framework, the material admits $M$ plastic events whose corresponding critical stresses are realized by sampling $P[\sigma]$, $M$ times --- the lowest of which is associated with the first plastic event. This stress value is is characterized by the stress scale, $\sigma^{*}_{1}$, given by~\cite{OrderStat,Bouchaud1997,Derlet2015a}
\begin{equation}
\frac{1}{M}=\int_{0}^{\sigma^{*}_{1}}d\sigma P[\sigma]=P_{<}[\sigma^{*}_{1}]. \label{Eqn1}
\end{equation}
Here $P_{<}[\sigma]$ is the cumulative distribution function of $P[\sigma]$. $M$ is expected to be linearly proportional to the deforming volume $L^{D}$, and via, $M=\rho L^{D}$, defines a density $\rho$ of plastic events the material can admit. Here $D$ is the spatial dimension of the deforming volume. Assuming that $P[\sigma]\sim\sigma^{(1-\gamma)/\gamma}$ as $\sigma$ approaches zero, for large enough $L$ (or $M$), Eqn.~\ref{Eqn1} limits to $M\times P[\sigma^{*}]\times d\sigma\sim M\times(\sigma^{*}_{1})^{(1-\gamma)/\gamma}\times\sigma^{*}_{1}\sim 1$ which gives
\begin{equation}
\sigma^{*}_{1}\sim \left(\frac{1}{M}\right)^{\gamma}\sim\left(\frac{1}{L^{D}}\right)^{\gamma}. \label{EqnPL}
\end{equation}
Such a power-law relation is expected for asymptotically large $M$. Ref.~\cite{Derlet2016} found Eqn.\ref{EqnPL} to hold for the stress at which the  first ``pop-in'' event occurs in a nano-indentation experiment, as a function of indenter volume, and also for the stress at which the first discrete plastic event occurs in one dimensional dislocation simulations. The goal of the present paper is to investigate the origin of this result for the latter one dimensional dislocation model.

Sec.~\ref{SecDD} outlines the one dimensional dislocation dynamics model used for the present study and sec.~\ref{SecRes} contains a detailed investigation of the origin of the statistical properties of the critical stress and plastic strain associated with the first plastic event. This is done by linearising the dislocation equations of motion and studying the properties of the resulting dynamical matrix as a function of applied external shear stress until the critical stress is reached. It is found the corresponding eigenmodes are spatially localized, indicating the eigenmode eventually responsible for the onset of the first plastic event always involves a low  number (usually one) of dislocations. For this eigenmode a simple relation exists between its eigenvalue prior to loading and the critical stress of the first plastic event, which may be understood within the framework of catastrophe theory. These two results demonstrate the statistics of the first critical stress are related to the statistics of the lower end of the eigenvalue spectrum of the dynamical matrix. By analysing the resulting irreversible plastic strain, it is found that in most cases the resulting plasticity spatially correlates with the structure of the unstable mode which triggers it. Sec.~\ref{SecDis} discusses these results in the context of eigenvalue perturbation theory which, via the fold catastrophe, directly relates the statistical properties of the dynamical matrix elements, its eigenvalues and that of the first critical stress into one unified picture. It concludes with a brief summary and discussion on the possible generality of these findings.

\section{Dislocation dynamics model} ~\label{SecDD}

A one dimensional configuration of $N$ infinitely long and straight edge dislocations is considered within the periodic interval $[0,d)$. Each dislocation (with Burgers vector along the line) interacts via a long range elastic interaction and also experiences a sinusoidal internal stress field of amplitude $\tau_{0}$ and period $\lambda_{0}$. In this model, the explicit dislocation content may be viewed as the mobile dislocation population, whilst the internal field represents a static mean-field description of the immobile dislocation population.  For further details see refs.~\cite{Derlet2013,Derlet2016}. $d$ will be referred to as the size of the system. The present work will consider a regime of large enough sizes such that a change of size will not fundamentally affect the nature of the plasticity, but only affect the statistics of how $d\rightarrow\infty$ bulk plasticity is probed. This viewpoint will break down for small enough $d$ where the choice of boundary conditions do matter --- a regime which is not investigated in the present work.
   
The (over-damped) equation of motion for the $i$th dislocation is given by
\begin{equation}
B\frac{dx_{i}}{dt}=F_{i}[\sigma]=\left(\sigma_{\mathrm{e}}\left[x_{i}\right]+\sigma\right)b+\sum_{j}f\left[x_{i}-x_{j}\right], \label{EqnEoM}
\end{equation}
where $b$ is the Burgers vector magnitude, $B$ is the damping coefficient, and $\sigma$ is the applied external stress. In the above, the static internal stress field is given by
\begin{equation}
\sigma_{\mathrm{e}}\left[x\right]=\tau_{0}\cos\left[\frac{2\pi x}{\lambda_{0}}\right],\label{EqnTauI}
\end{equation}
and the force between two dislocations seperated by a distance $x$ is given by
\begin{equation}
f[x]=\frac{Gb^{2}}{2\pi(1-\nu)}\times\frac{\pi}{d}\cot\left[\frac{\pi x}{d}\right].\label{EqnDDInt}
\end{equation}
In the above, $G$ is a shear modulus and $\nu$ is Poisson's ratio. Eqn.~\ref{EqnDDInt} is based on the isotropic elastic interaction between two edge dislocations of infinite length with parallel line directions, and has already been lattice summed (see appendix of ref.~\cite{Derlet2013})--- it therefore has a range of $d/2$. In the limit of $d\rightarrow\infty$, Eqn.~\ref{EqnDDInt} limits to $\sim1/x$.

The present work considers how the dislocation configuration responds to an increasing external stress $\sigma$ via the static condition $F_{i}[\sigma]=0$ for $i=1$ to $N$. As a numerical simulation, this entails Eqn.~\ref{EqnEoM} is numerically evolved until all forces on the dislocation network are zero, for each small increment of $\sigma$. The resulting configuration is then identified with the current external stress value $\sigma$. The resulting plastic strain associated with this relaxation is defined as
\begin{equation}
\delta\varepsilon=\frac{b}{dh}\sum_{i}\delta x_{i}, \label{EqnPStrain}
\end{equation}
which is based on a simplified Elshelby picture of the far-field strain signature due to a localized plastic event~\cite{Eshelby1957}. In the above, $h$ is an arbitrary length representing the width of the one dimensional medium in which the dislocations exist, with the product $hd$ being its two dimensional volume of the deforming material --- see ref.~\cite{Derlet2013}. Note that for this model, both the external and internal stresses, as well as the resulting plastic strain have only a pure shear component. 

Following refs.~\cite{Derlet2013,Derlet2016}, the initial dislocation configuration at zero loading is produced by placing $N$ dislocations at random positions within the $[0,d)$ interval and relaxing the system to its zero-force configuration. This then will be referred to as the initial configuration prior to loading.

\section{Results} \label{SecRes}

 \subsection{Plastic evolution to the first discrete plastic event} \label{SSecPE}
 
\begin{figure}
\includegraphics[clip,width=0.47\textwidth]{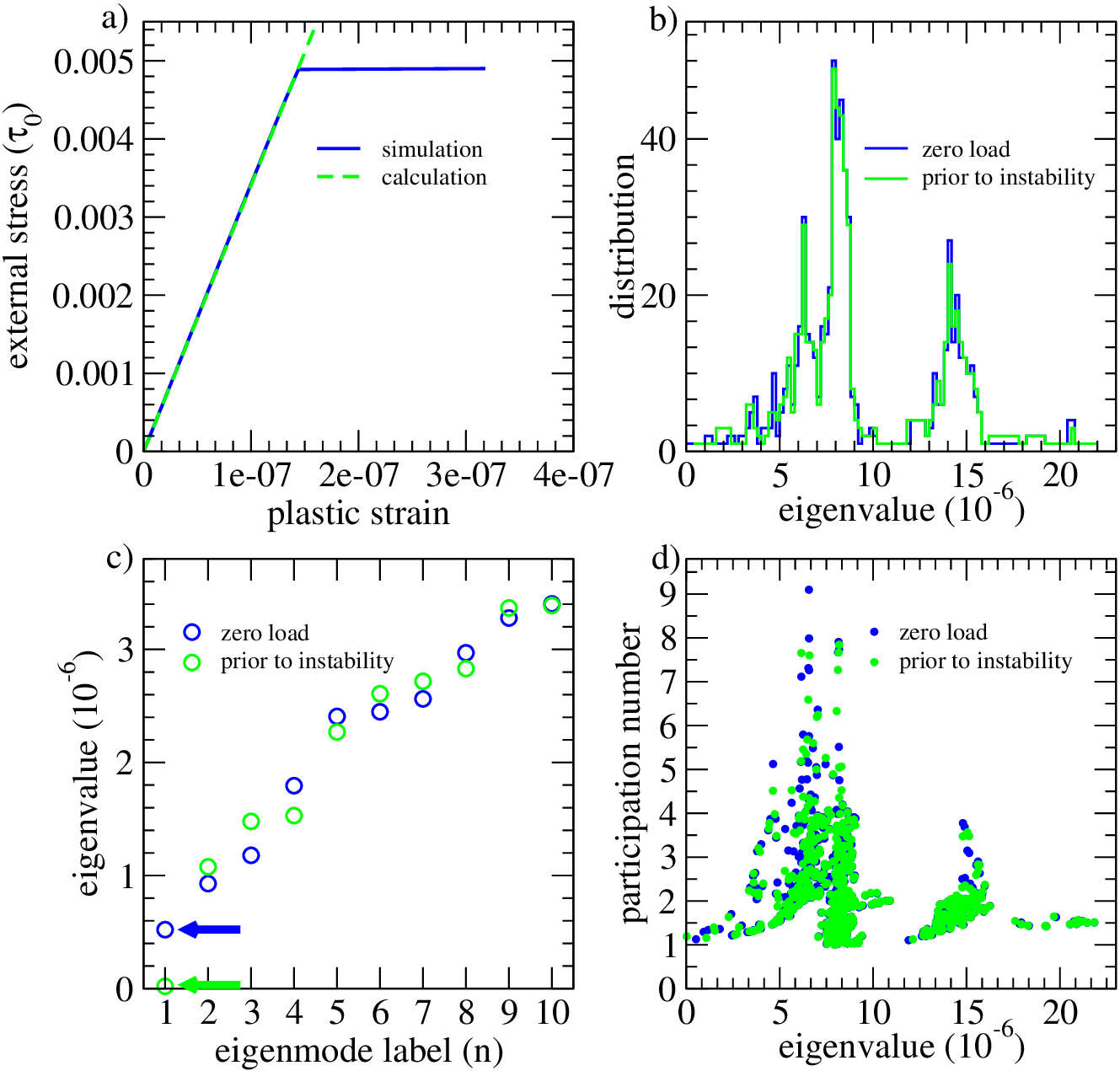}
\caption{a) Plot of numerically simulated stress versus plastic strain evolution, beginning at zero load and ending with the arrest of the first discrete plastic strain event. Also shown is the prediction arising from the linearization of the dislocation equations of motions: Eqn.~\ref{EqnLPSEV}. b) Distribution of eigenvalues and c) plot of the ten lowest eigenvalues of the dynamical matrix corresponding to the dislocation configuration prior to loading and prior to the plastic instability, with the lowest eigenvalue approaching zero as the instability is neared (see arrowed data in c). d) Participation number (Eqn.~\ref{EqnPN}) of corresponding eigenvectors showing the approximate number of dislocations involved in each eigenmode. \label{FigSim}}
\end{figure}

 In what follows, a system of size $d=1280$ $\mu$m is considered with $N=640$ dislocations, $\tau_{0}=10\mathrm{MPa}$ and  $\lambda_{0}=2\mu\mathrm{m}$. Following, ref.~\cite{Derlet2013}, this can be viewed as representing a Cu crystal ($b=2.55$ \AA) with a dilute immobile and mobile dislocation content. The stress versus plastic strain curve, up to and including the first plastic event, for a particular initial dislocation configuration is displayed in Fig.~\ref{FigSim}a. The plastic evolution is characterized by an initial linear and reversible regime of plasticity followed by a strongly non-linear regime prior to the first discrete (and irreversible) plastic event --- the plateau region of the curve The figure shows the stress versus plastic strain evolution until arrest of this first irreversible plastic event. Such a response is typical, with the gradient of the linear regime varying little with respect to the starting dislocation configuration.
 
 The linear regime prior to the instability, seen in Fig.~\ref{FigSim}a, suggests a linearization of Eqn.~\ref{EqnEoM} should describe the initial linear plastic response. By writing the initial zero load dislocation configuration as $\{x^{0}_{i}\}$, and its response to an external stress as $\{x_{i}=x^{0}_{i}+u_{i}\}$, the linearization of  Eqn.~\ref{EqnEoM} at zero force is
\begin{equation}
-\sum_{j}\Lambda_{ij}u_{j}+\sigma b=0, \label{EqnLEoM}
\end{equation}
where the dynamical matrix, $\Lambda_{ij}$, is equal to
 \begin{multline}
\Lambda_{ij}[\{x^{0}_{i}\}]=\\-\left(\sigma'_{\mathrm{e}}[x^{0}_{i}]b-\sum_{k}f'\left[x^{0}_{i}-x^{0}_{k}\right]\right)\delta_{i,j}-f'\left[x^{0}_{i}-x^{0}_{j}\right]. \label{EqnDM}
 \end{multline}
In the above and throughout the present work, the summations over dislocation number span from 1 to $N$. The solution to Eqn.~\ref{EqnLEoM} is
\begin{equation}
u_{i}=\sigma b\sum_{j}\left[\tilde{\Lambda}^{-1}\right]_{ij}=\sigma b\sum_{n}\frac{\sum_{j}\left[\mathbf{u}_{n}\right]_{j}}{e_{n}}\left[\mathbf{u}_{n}\right]_{i}, \label{EqnLSolnEV}
\end{equation}
giving the plastic strain (Eqn.~\ref{EqnPStrain}) as
\begin{equation}
\varepsilon=\frac{\sigma b^{2}}{dh}\sum_{ij}\left[\tilde{\Lambda}^{-1}\right]_{ij}=\frac{\sigma b^{2}}{dh}\sum_{n}\frac{\left(\sum_{j}\left[\mathbf{u}_{n}\right]_{j}\right)^{2}}{e_{n}}, \label{EqnLPSEV}
\end{equation}
where the last equalities are written in terms of the eigenvalues and normalized eigenvectors of $\tilde{\Lambda}$, defined as $\tilde{\Lambda}\mathbf{u}_{n}=e_{n}\mathbf{u}_{n}$. It is noted that the eigenvalues have units of stress. Using the zero load configuration of Fig.~\ref{FigSim}a, the prediction of Eqn.~\ref{EqnLPSEV} is also plotted, showing good agreement for the linear plastic regime prior to the first discrete plastic event.

The energy, from which Eqn.~\ref{EqnLEoM} is obtained, may be written as
\begin{equation}
E=\frac{1}{2}\sum_{ij}u_{i}\Lambda_{ij}u_{j}-\sigma b\sum_{i}u_{i}. \label{EqnLEnergy}
\end{equation}
Substitution of Eqn.~\ref{EqnLSolnEV} into the above gives
\begin{align}
E=&-\frac{1}{2}(\sigma b)^2\sum_{n}\frac{\left(\sum_{j}\left[\mathbf{u}_{n}\right]_{j}\right)^{2}}{e_{n}}\\
=&-\frac{1}{2}\times hd\times \sigma\times \varepsilon[\sigma]\\ 
\dot{=}&-\frac{1}{2}\times hd\times \sigma^{2}\times \frac{1}{G_{\mathrm{ael}}}, \label{EqnLEnergy1c}
\end{align}
where 
\begin{equation}
\frac{1}{G_{\mathrm{ael}}}=\frac{b^{2}}{dh}\sum_{n}\frac{\left(\sum_{j}\left[\mathbf{u}_{n}\right]_{j}\right)^{2}}{e_{n}} \label{EqnAESM}
\end{equation}
may be viewed as an anelastic shear modulus. Eqn.~\ref{EqnLEnergy1c} can be best understood by adding the energy due to elastic shear deformation, characterized by the elastic shear modulus $G$, resulting in a deformation energy depending quadratically on the applied external stress:
\begin{equation}
E_{\mathrm{{def}}}=\frac{1}{2}\times hd\times \sigma^{2}\times\left(\frac{1}{G}-\frac{1}{G_{\mathrm{ael}}}\right). \label{EqnLEnergy1d}
\end{equation}
Eqn.~\ref{EqnLEnergy1d} embodies a simple theory of linear anelasticity in which reversible plastic deformation results in a reduced effective elastic stiffness. The above constitutes a mathematical description of a regime of plastic activity recently studied in the context of quasi-reversible deformation occuring between regimes of plasticity characterized by universal avalanche phenomenon~\cite{Szabo2015}.

A similar linearlization procedure can be performed at any step along the deformation curve by taking the locally relaxed coordinates from the numerical solution to Eqn.~\ref{EqnEoM}, at a given applied external stress, to give $\tilde{\Lambda}[\{x^{\sigma}_{i}\}]$. Indeed, the proportionality factor between plastic strain and the stress perturbation in Eqn.~\ref{EqnLPSEV} gives directly the inverse of the gradient of the simulated plastic deformation curve at the chosen applied external stress.

Fig.~\ref{FigSim}b plots the distribution of eigenvalues obtained from the dynamical matrix evaluated for the dislocation configuration prior to loading and just before the occurrence of the discrete plastic event. The distributions show only minor overall differences for the plastic evolution to the instability. Fig.~\ref{FigSim}c plots the corresponding low-end regime of eigenvalues as a function of eigenmode label, showing the very lowest eigenvalue approaches zero close to the instability, with the other eigenvalues either increasing or decreasing with respect to their initial value. This trend is found to be generally true and results in Eqns.~\ref{EqnLSolnEV} and \ref{EqnLPSEV} diverging at the instability, indicating a breakdown of the linearization approach and a divergent gradient of plastic strain with respect to applied external stress.

Eqn.~\ref{EqnLSolnEV} reveals the displacement vector $\mathbf{u}$, in which the $i$th element represents the displacement of the $i$th dislocation, arises from a weighted sum of the eigenmodes. Since the weight for the $n$th eigenmode is inversely proportional to its eigenvalue, as the instability is approached, the displacement vector is increasingly dominated by the eigenmode with the lowest eigenvalue. Inspection of the non-vansishing elements of this eigenmode therefore gives insight into those dislocations responsible for the instability.

To obtain an estimate of the number of dislocations involved in an eigenmode, the participation number~\cite{Bell1970} is used:
\begin{equation}
\mathrm{PN}_{n}=\left(\sum_{i}\left[\mathbf{u}_{n}\right]_{i}^{4}\right)^{-1},\label{EqnPN}
\end{equation}
which gives a value of unity when all but one element of the normalized eigenvector is equal to zero, and a value of $N$ when each element has a value of $1/\sqrt{N}$. Fig.~\ref{FigSim}d plots the participation number of all eigenvectors for the dynamical matrix evaluated at both zero load and at a stress just prior to the instability. The figure demonstrates only a few dislocations are involved in each eigenvector and those with the very lowest eigenvalues involve approximately one dislocation. Thus, in this case, the unstable mode involves only one dislocation as the trigger to subsequent plastic activity. This result is found to be generally true.

\begin{figure}
 	\includegraphics[clip,width=0.47\textwidth]{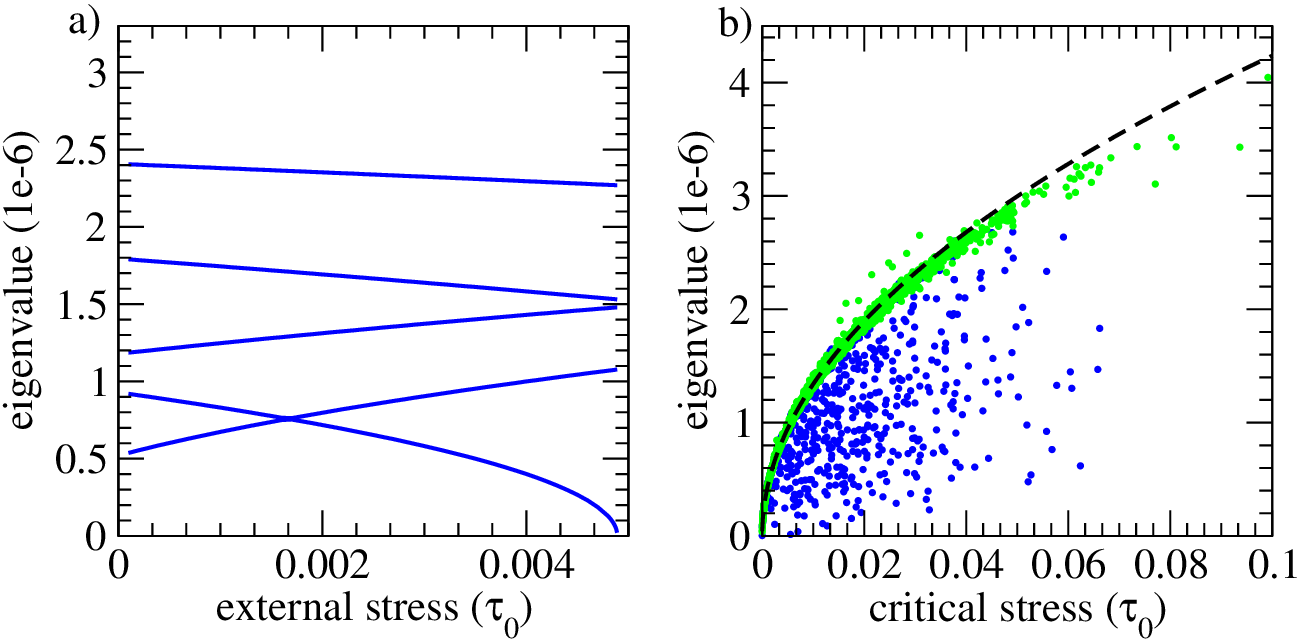}
 	\caption{a) Plot of the applied external stress dependence of the six lowest eigenvalues. b) Scatter plot of lowest (blue) eigenvalue prior to loading versus the critical stress of the first plastic event, derived from 1000 loading simulations. In b) a scatter plot of the eigenvalue corresponding to the initial eigenmode with the highest overlap to the unstable eigenmode is shown in green.\label{FigEVal}}
 \end{figure}

 Fig.~\ref{FigEVal}a displays the six lowest eigenvalues as a function of applied stress derived from diagonalization of $\tilde{\Lambda}[\{x^{\sigma}_{i}\}]$. For this particular starting configuration, it is seen that the mode associated with the second lowest eigenvalue prior to loading eventually leads to the instability of the first discrete plastic event. Thus the two lowest eigenvalues of the initial and unstable dynamical matrices in Fig.~\ref{FigSim}c have interchanged eigenmodes.  
  
 The functional form of the eigenvalue's external stress dependence (in Fig.~\ref{FigEVal}a) for the mode associated with the instability is very well described by the functional form
\begin{equation}
e_{*}[\sigma]=e_{*}[0]\sqrt{1-\frac{\sigma}{\sigma_{\mathrm{c}}}}, \label{EqnRR}
\end{equation}
suggesting a fold catastrophe~\cite{Thom1975,Arnold1992} is at play --- a feature that has also been seen in multi-dimensional potential energy landscapes of glassy materials~\cite{Wales2001,Maloney2006}. 

To investigate the generality of this result one thousand loading simulations to the first plastic event were performed and for each, the lowest few eigenvalues were determined as a function of external stress. Fig.~\ref{FigEVal}b displays the scatter plot between the lowest eigenvalue at zero loading versus its associated critical stress. Writing Eqn.~\ref{EqnRR} as $e_{*}[\sigma]=c\sqrt{\sigma_{\mathrm{c}}-\sigma}$, demonstrates if the fold catastrophe relation would hold, then the upper bounds of such a scatter plot should correspond to $e_{*}[0]=c\sqrt{\sigma_{\mathrm{c}}}$. That is, if the lowest eigenvalue corresponds to the unstable mode then it must be on the upper bound envelope and if it is not, it must be below it. Inspection of the figure, which also plots $e_{*}[0]=c\sqrt{\sigma_{\mathrm{c}}}$ versus $\sigma_{\mathrm{c}}$ for $c$ determined from the configuration of Fig.~\ref{FigEVal}a, clearly shows that such a functional form does hold. A somewhat more direct approach is to plot the eigenvalue corresponding to the initial eigenmode with the highest overlap with the unstable eigenmode. Fig.~\ref{FigEVal}b plots this data (in green) and again demonstrates the close adherence to the fold catastrophe relation. The origin of the outlier points above and below the fold catastrophe curve will be discussed respectively in secs.~\ref{SSecCT} and \ref{SSecPS}.

\subsection{Origin of the eigenvalue spectrum of the dynamical matrix} \label{SubSecEVS}

The current results are related to the way a starting configuration is selected --- $N$ dislocations are randomly placed within the interval $[0,d)$ and relaxed to a local energy minimum. The energy of this disordered dislocation configuration is higher than the global energy minimum for the current model and thus the starting configuration can be viewed as being out of equilibrium. The rational for this choice is that the mobile dislocation content of a material system is not expected to be close to equilibrium~\cite{Zaiser2006}, since it forms only a small part of the total dislocation content --- the other part being that of the immobile dislocation content (which in the current model is represented via Eqn.~\ref{EqnTauI}). Whether or not the total dislocation content can be understood within an equilibrium or out-of-equilibrium framework remains a topic of continued debate~\cite{Zaiser2006,DisSolids}. 
 
 \begin{figure}
 	\includegraphics[clip,width=0.47\textwidth]{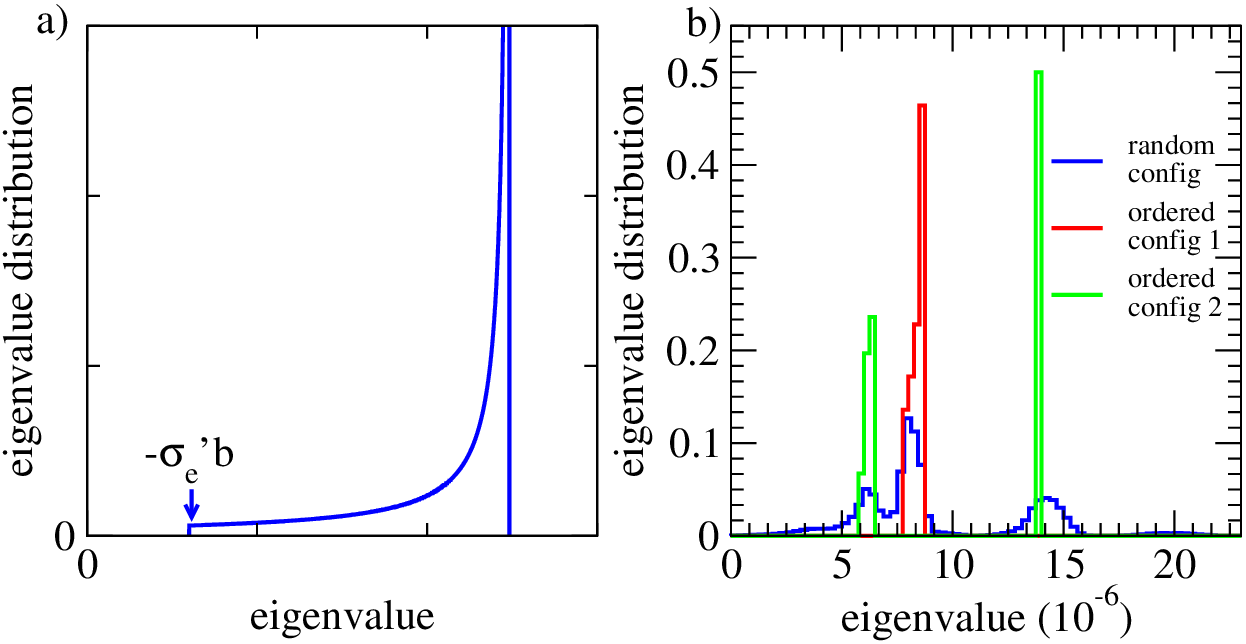}
 	\caption{a) Analytical DOS arising from an ordered array of dislocations with lattice constant $\lambda_{0}$ and $d=\infty$. b) Numerical DOS for the $d=1028$ $\mu$m system for an ordered array with lattice constant $\lambda_{0}$ (configuration 1) and $2\lambda_{0}$ (configuration 2 --- see text) with the DOS of Fig.~\ref{FigSE}c.\label{FigDis}}
 \end{figure}
   
For the current model and choice of its parameters (in particular $N=d/\lambda_{0}$) the global energy minimum is an ordered array of $N$ dislocations with a ``lattice constant'' equal to $\lambda_{0}$. This configuration, and variations from it give insight into both the form of the eigenvalue spectrum shown in Fig.~\ref{FigSE}b and some aspects of deformation. Because of the discrete translation symmetry of this configuration, the corresponding linearized matrix for $d\rightarrow\infty$ may be exactly diagonalized, giving the eigenvalue,
\begin{equation}
e_{k}=\sigma'_{\mathrm{e}}b+A\left(\frac{\pi^{2}}{3}-\left(\mathrm{Li}_{2}\left[e^{ik\lambda_{0}}\right]+\mathrm{Li}_{2}\left[e^{-ik\lambda_{0}}\right]\right)\right), \label{EqnEvalOrdered}
\end{equation}
with the label, $k$, being a one dimensional reciprocal space vector whose value is between $\pm\pi/\lambda_{0}$. The corresponding eigenvector is $[\mathbf{u}_{k}]_{j}\sim\exp(ik\lambda_{0} j)$ where $j$ labels the $j$th dislocation. In the above $A=Gb^2/(2\pi(1-\nu)\lambda_{0})$, $\mathrm{Li}_{2}\left[\cdot\right]$ is the poly-logarithm (Jonqui\`ere's function) of order 2, and $\sigma'_{\mathrm{e}}=\sigma'_{\mathrm{e}}[x_{j}]$. This latter equality is independent of $j$ since the dislocation within each $\lambda_{0}$ interval is at the same location. The lowest eigenvalue corresponds to the case of $k=0$ in which the ordered array responds uniformly to the applied stress: $ e_{k}=\sigma'_{\mathrm{e}}b$.   The equations of motion, Eqn.~\ref{EqnEoM}, then simplify to $\sigma_{\mathrm{e}}\left[x_{j}\right]=-\sigma$, giving $x_{j}=\lambda_{0}(j+\arccos [-\sigma/\tau_{0}]/(2\pi))$. 

Thus for the lowest eigenvalue, Eqn.~\ref{EqnEvalOrdered} becomes
\begin{equation}
e_{k=0}=\frac{\tau_{0}2\pi b}{\lambda_{0}}\sqrt{1-\left(\frac{\sigma}{\tau_{0}}\right)^{2}},\label{EqnEvalOrdered1}
\end{equation}
which limits to
\begin{equation} 
\frac{\tau_{0}2\pi b\sqrt{2}}{\lambda_{0}}\sqrt{1-\frac{\sigma}{\tau_{0}}} 
\end{equation}
and a simple fold catastrophe as $\sigma\rightarrow\tau_{0}$.

The analytical eigenvalue distribution of Eqn.~\ref{EqnEvalOrdered} is schematically shown in Fig.~\ref{FigDis}a, and also evaluated numerically for the $d=1280$ $\mu$m system in Fig.~\ref{FigDis}b (shown in red), along with the eigenvalue spectrum characteristic of the initial dislocation configurations used in sec.~\ref{SecRes}. Fig.~\ref{FigDis}b therefore demonstrates the second peak of this latter distribution arises from finite segments of ordered dislocations, which the participation number data of Fig.~\ref{FigSim}d show to involve up to eight dislocations.

The above approach can be extended to consider the case of two dislocations within every second $\lambda_{0}$ interval. The corresponding eigenvalue spectrum is shown in Fig.~\ref{FigDis}b (shown in green), giving rise to two peaks, one of which may be viewed as an ``acoustic branch'' and the other (higher eigenvalue) peak as an ``optical branch'' since in this case the unit cell is of length $2\lambda_{0}$. Again there is a good correspondence to two peaks of the eigenvalue spectrum characteristic of the initial dislocation configurations. Other ordered configurations may be used to gain understanding of the remaining peaks.

\subsection{Size effects in critical stress and catastrophe theory} \label{SSecCT}

\begin{figure}
\includegraphics[clip,width=0.47\textwidth]{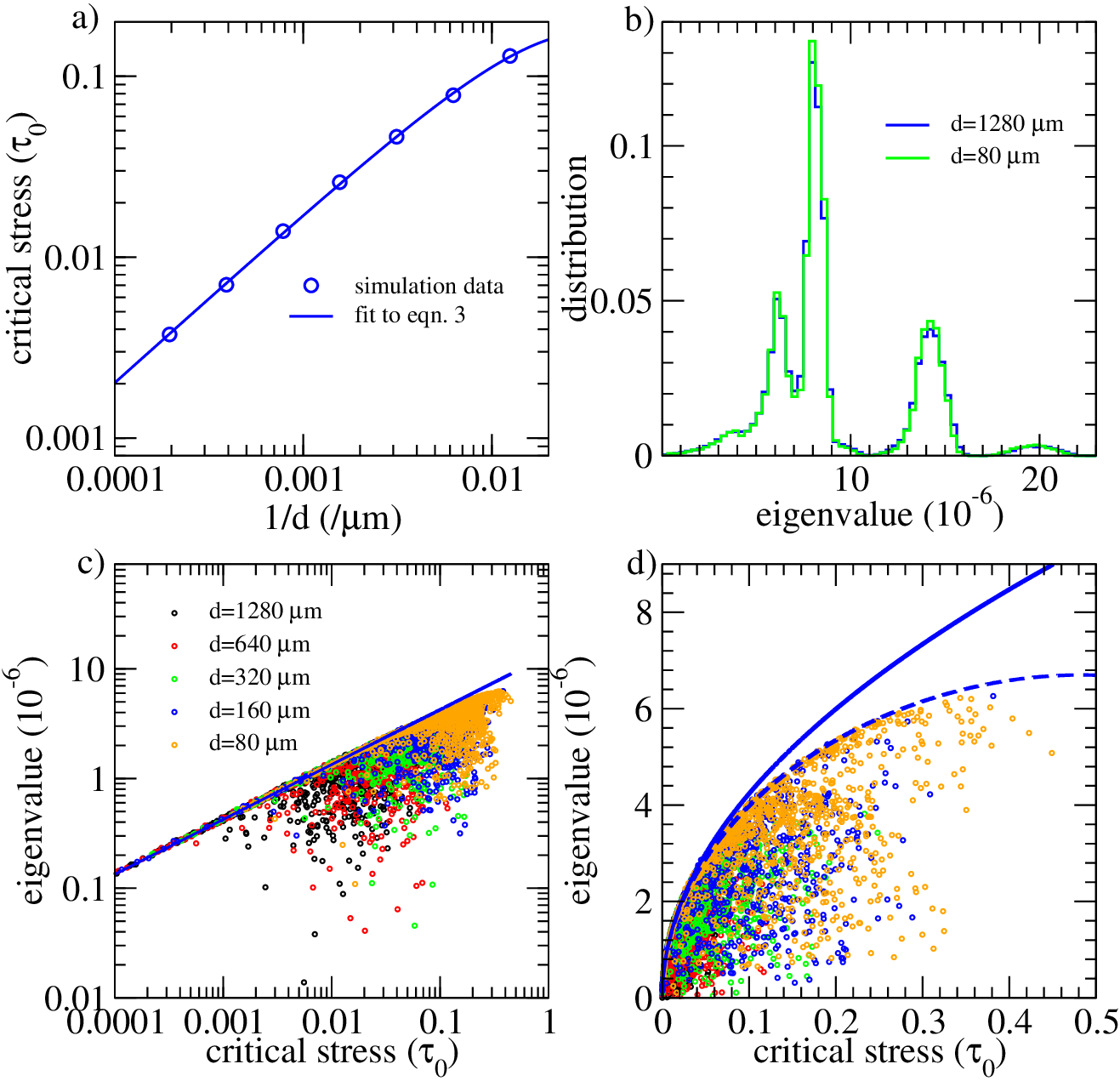}
\caption{a) Plot of the average critical stress versus the inverse system size and it's fit using an exponentially truncated power law, b) The distribution of eigenvalues derived from the dynamical matrix of the linearized dislocation equations of motion for the largest and smallest considered system sizes. c) Log-log and d) linear scatter plot of lowest eigenvalues prior to loading versus the critical stress of the first plastic event for the considered system sizes. In both plots the solid line represents the prediction due to catastrophe theory for a fold instability, whereas the dashed line in d) shows the prediction arising from an additional higher order term of the potential energy landscape giving the $O(5)$ fold.\label{FigSE}}
\end{figure}

To investigate how the critical stress and associated eigenvalue spectrum might depend on system size, ten thousand loading simulations are performed for $d=640$, $320$, $160$ and $80$ $\mu$m for the same dislocation density as the $d=1280$ $\mu$m system studied so far. Data for $d=2560$ and $5120$ $\mu$m is also shown. Fig.~\ref{FigSE}a plots the mean critical stress versus the inverse of the system length along with a fit to an exponentially truncated power law: $\exp(-l/d)/d^{\gamma}$, giving an exponent of $\gamma\approx0.9$. The value of $\gamma$ differs slightly from that of ref.~\cite{Derlet2016}, which derived its parameters using a fixed (high) strain rate loading mode. The value of the exponent $\gamma$ gives a low-stress power-law exponent for the master critical stress distribution equal to $(1-\gamma)/\gamma\approx0.11$. Ref.~\cite{Derlet2016} found that the weak exponential truncation can arise from deviations away from the Weibull limit of asymptotically large $M$.

Fig.~\ref{FigSE}b displays the average eigenvalue spectrum of the dynamical matrix associated with the dislocation configuration prior to loading, for the case of $d=1280$ $\mu$m and $d=80$ $\mu$m, and demonstrates that overall it changes little for the system sizes so far considered suggesting there is little fundamental difference in the nature of the first plastic event for system sizes down to $d=80$ $\mu$m. 

Fig.~\ref{FigSE}c displays a log-log scatter plot of the lowest eigenvalue versus the associated critical stress for all considered systems sizes less than $d=1024$ $\mu$m. Again, a distinct upper bound limit clearly exists and for small enough critical stresses it follows Eqn.~\ref{EqnRR}, which is also plotted in the figure. For larger stresses, the relation between the unstable eigenvalue and the critical stress appears to deviate from this fold catastrophe form --- a trend that is best seen in the linear plot of the same data, Fig.~\ref{FigSE}d. To address this aspect, the details of catastrophe theory must be considered.

Catastrophe theory classifies the way in which a potential energy landscape approaches an instability~\cite{Thom1975,Arnold1992,Wales2001}. The simplest of these is the so-called fold catastrophe~\cite{Wales2001,Maloney2006}, whose potential energy landscape as the instability is approached takes the form:
\begin{equation}
	E^{(3)}=\frac{1}{3}Ax^{3}-Bx\delta. \label{EqnFold3}
\end{equation}
Here $x$ parametrizes the last remaining relevant coordinate and $\delta$ is the external stimulus, which in our case is $\sigma-\sigma_{\mathrm{c}}$. The minimum and maximum stationary points, this equation characterizes, are at $x^{(3)}_{\pm}=\pm\sqrt{B\delta/A}$ with the corresponding curvatures (eigenvalues) equalling $e^{(3)}_{\pm}=\pm2\sqrt{AB\delta}$. The barrier energy at finite $\delta$ then becomes $\Delta U^{(3)}=2\sqrt{2(B\delta)^{3}/(9A)}$. The so-called fold ratio $6\Delta U^{(3)}/(e^{(3)}_{-}(2x^{(3)}_{-})^{2})$ is equal to unity when these relations are maintained. $e^{(3)}_{-}$ gives directly Eqn.~\ref{EqnRR}.
These relations are formally only valid in the limit of the $\sigma\rightarrow\sigma_{\mathrm{c}}$. The present work shows however that for $d=1280$ $\mu$m, the fold catastrophe relation is followed very well for the stress interval $[0,\sigma_{\mathrm{c}})$. The applicability of the fold relations for non-negligible finite values of $\sigma-\sigma_{\mathrm{c}}=\delta$ has also been seen in atomistic simulations of stressed model amorphous solids~\cite{Maloney2006}, for which the material instability is an atomic scale structural transformation.

\begin{figure}
\includegraphics[clip,width=0.47\textwidth]{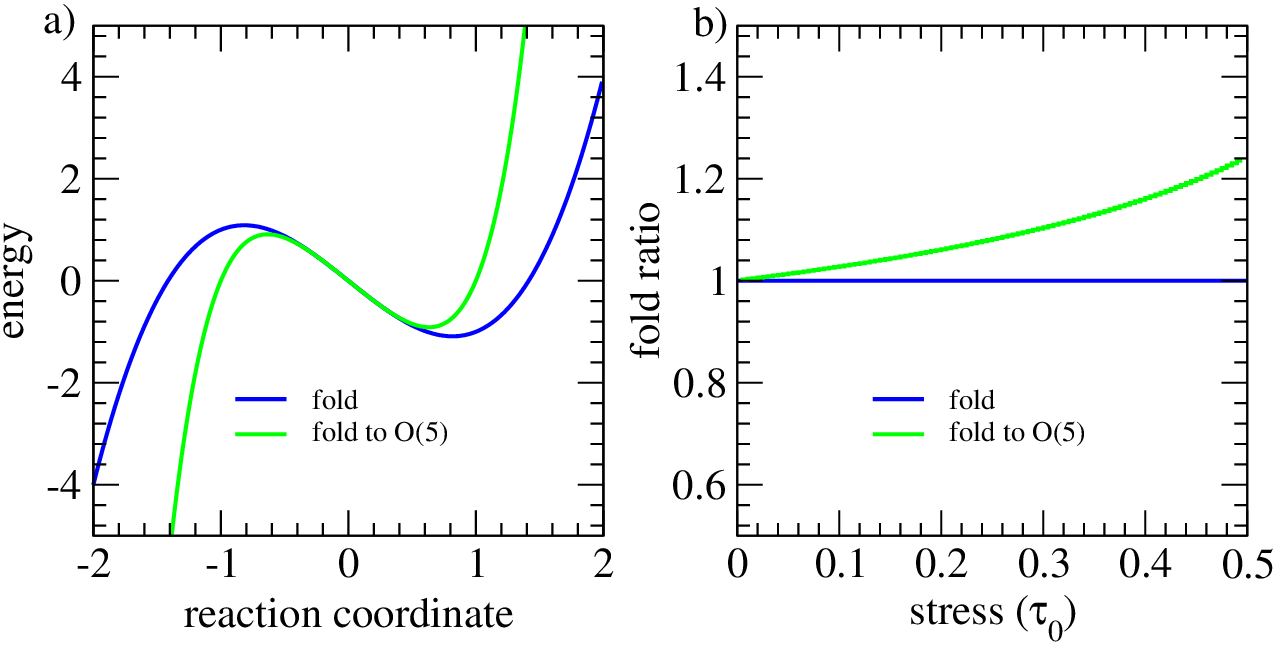}
\caption{a) Plot of relevant potential energy landscape close to a fold catastrophe (Eqn.~\ref{EqnFold3}). Also shown is the $O(5)$ order fold catastrophe energy (Eqn.~\ref{EqnFold5}), b) The fold ratio as a function of the external stress for the $O(5)$ fold. \label{FigFold}}
\end{figure}

To not fundamentally change the catastrophe from that of a fold, the next highest term to be added will involve the 5th power:
\begin{equation}
E^{(5)}=\frac{1}{3}Ax^{3}-Bx\delta+\frac{1}{5}Cx^{5}, \label{EqnFold5}
\end{equation}
whose minimum and maximum stationary points are given by
\begin{equation}
x^{(5)}_{\pm}=\pm\sqrt{\frac{\left(A^2+4 B C \delta\right)^{1/2}-A}{2C}},
\end{equation}
with corresponding eigenvalues:
\begin{equation}
e^{(5)}_{\pm}=\pm\sqrt{\frac{2(A^2+4 B C \delta)\left(\left(A^2+4 B C \delta\right)^{1/2}-A\right)}{C}}.
\end{equation}
This will be referred to as the $O(5)$ fold catastrophe. The above may be written in terms of $e^{(3)}_{+}$ giving
\begin{equation}
e^{(5)}_{+}[\sigma]=\sqrt{\frac{2 \left(D e^{(3)}_{+}[\sigma]^2+1\right) \left(\left(D  e^{(3)}_{+}[\sigma]^2+1\right)^{1/2}-1\right)}{D}}, \label{EqnEval5}
\end{equation}
where $D=C/A^{3}$ is the remaining free parameter of Eqn.~\ref{EqnFold5}. For an appropriate choice of $D$, Eqn.~\ref{EqnEval5} evaluated at $\sigma=0$ is able to describe the relationship between critical stress and corresponding unstable mode eigenvalue for the entire considered applied external stress range --- see Fig.~\ref{FigSE}d, which plots Eqn.~\ref{EqnEval5} at $\sigma=\sigma_{\mathrm{c}}$ as a dashed curve. Here $D$ (and therefore $C$) must be negative reflecting the fact that the maximum of Eqn.~\ref{EqnEval5} occurs at a stress $\sigma_{\mathrm{max}}$ defined by $e^{(3)}_{+}[\sigma_{\mathrm{max}}]=\sqrt{-5/(9D)}$, and is equal to $\sqrt{-8/(27D)}$.

Fig.~\ref{FigFold}a plots the resulting potential energy landscape for both the fold and $O(5)$ fold catastrophe, demonstrating that such a 5th order term in Eqn.~\ref{EqnFold5} does not fundamentally change the nature of the instability, only changing the position and curvature of the local minimum and corresponding saddle-point. Inspection of Fig.~\ref{FigFold}b, shows however that due to these modified stationary points, the fold ratio~\cite{Wales2001}, which is identically one for the usual fold catastrophe, can deviate by up to 20\% for finite applied external stresses.

\subsection{Structure of the dynamical matrix} \label{SSecDM}

\begin{figure}
	\includegraphics[clip,width=0.47\textwidth]{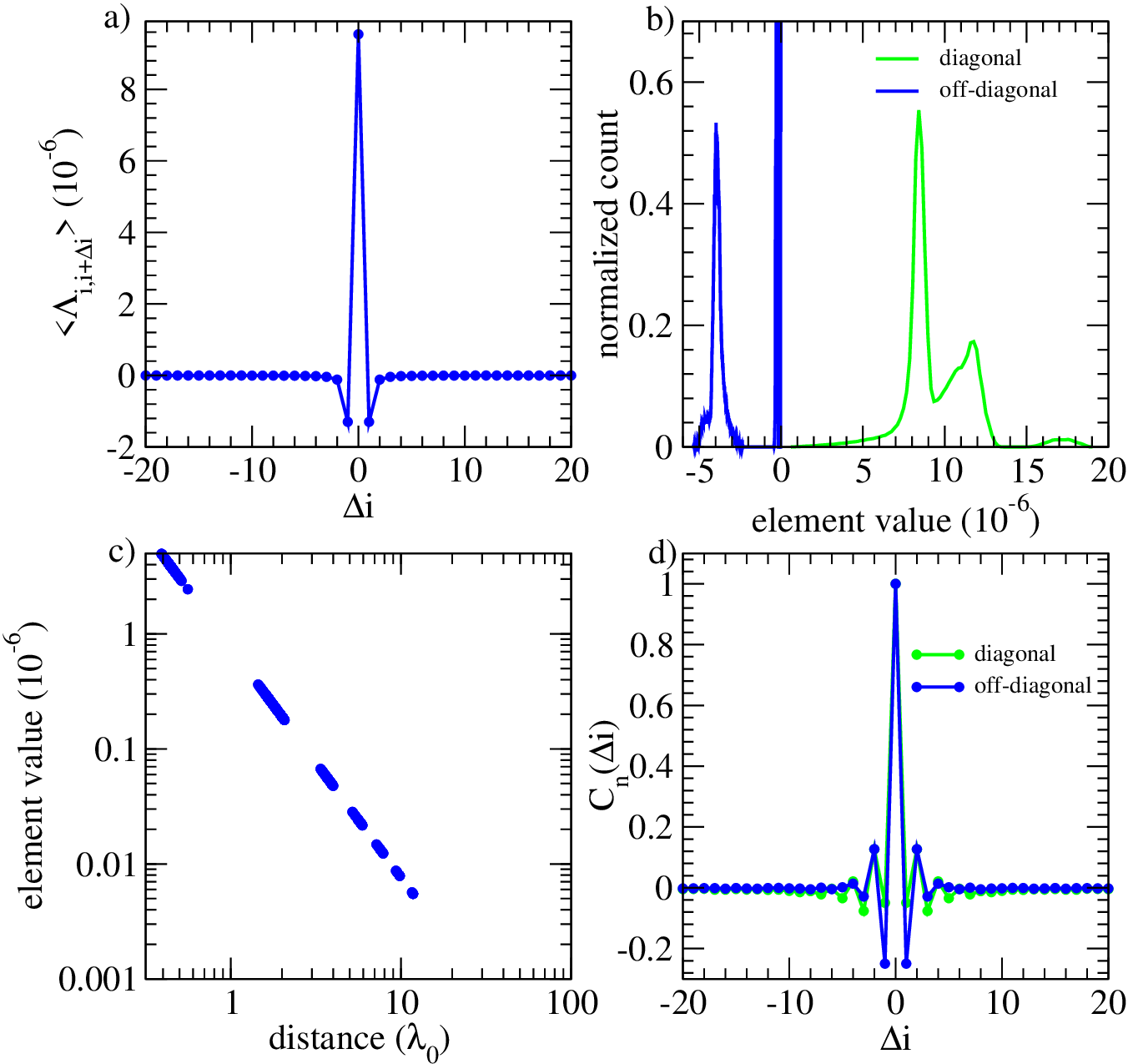}
	\caption{a) Average $i,i+\Delta i$ element of linearized matrix indicating a matrix which is predominantly tri-diagonal. b) Normalized histograms of diagonal and off-diagonal bands. c) Plot of $ij$th off-diagonal element value versus $x_{i}-x_{j}$. d) Intra-diagonal and intra-off-diagonal correlation functions. \label{FigDis1}}
\end{figure}

Further insight may be gained by inspecting the statistical properties of the dynamical matrix $\Lambda_{ij}$. In what follows, the labelling refers to a sequential list of dislocations with increasing initial position. Thus the nearest neighbour dislocations of the $i$th dislocation, barring periodic boundary conditions, have the labels $i-1$ and $i+1$. All statistics are derived from the $d=1280$ $\mu$m system size. Fig.~\ref{FigDis1}a plots the mean of $\Lambda_{i,i+\Delta i}$ and reveals that $\Lambda_{ij}$ is predominantly tri-diagonal with positive diagonal elements and negative off-diagonal elements. Fig.~\ref{FigDis1}b plots normalized histograms of these diagonal and off-diagonal band elements.

From Eqn.~\ref{EqnDM} there exist two contributions to each diagonal element of the dynamical matrix, the contribution arising from the elastic interaction between dislocations and the contribution arising from the internal static stress field. For the single slip system considered here, the former will always contribute positively to the diagonal element due to the mutual repulsion between dislocations, whereas the latter can either be positive or negative depending on the location of the corresponding dislocation. Detailed inspection reveals two classes of environments which lead to two distinct dislocation contributions, one arising from  regions in which there is more than one dislocation within a $\lambda_{0}$ interval and the other arising from regions in which neighbouring dislocations are spaced one per $\lambda_{0}$ interval. This latter contribution is responsible for the first diagonal element peak in (the green curve in Fig.~\ref{FigDis1}b) whose location is similar to that for the ordered configuration (the red curve shown in Fig.~\ref{FigDis}b). Such segments of order do not contribute to the low-end tail of the diagonal elements of the dynamical matrix. The second peak for the diagonal elements in Fig.~\ref{FigDis1}b arises from dislocation arrangements within a single $\lambda_{0}$ with its low-end tail being responsible for the low-end tail of the diagonal element distribution. This latter tail is due to the different locations the dislocations can take because of the static internal stress field and the surrounding dislocation structure. Inspection of the spatial structure of the initial dislocation configurations show that up to three dislocations can exist within a $\lambda_{0}$ interval.

Inspection of the negative off-diagonal elements show approximately 60\% of the elements are non-negligible and centred around a value whose magnitude is approximately half that of the diagonal elements. When this is not the case, the value is at least an order of magnitude smaller. The origin of this off-diagonal structure is best seen by inspection of $\Lambda_{ij}$ and the distance separating the dislocation pair $x_{i}-x_{j}$ --- Fig.~\ref{FigDis1}c. It is seen a large off-diagonal band element arises when a neighbouring (or next nearest neighbouring) dislocation is within the same $\lambda_{0}$ interval, whereas the smaller elements arise from neighbouring dislocations at more distant $\lambda_{0}$ intervals. The exponent of two in Fig.~\ref{FigDis1}c is related to the leading order, $1/r^{2}$, scaling of the off-diagonal element (Eqn.~\ref{EqnDM}).

Fig.~\ref{FigDis1}d plots the spatial correlation functions along the diagonal and the off-diagonal bands. Here
\begin{equation}
C^{n}_{\Delta i}=\frac{\left\langle\left(\Lambda_{i,i+n}-\overline{\Lambda}_{i,i+n}\right)\left(\Lambda_{i+\Delta i,i+\Delta i+n}-\overline{\Lambda}_{i+\Delta i,i+\Delta i+n}\right)\right\rangle_{i}}{\left\langle\left(\Lambda_{i,i+n}-\overline{\Lambda}_{i,i+n}\right)\left(\Lambda_{i,i+n}-\overline{\Lambda}_{i,i+n}\right)\right\rangle_{i}},
\end{equation}
where $n=0$ for the diagonal and $n=1$ for the off-diagonal bands. In the above, $\overline{\Lambda}_{ij}$, equals the mean $ij$th element and the average, $\langle\cdot\rangle_{i}$, is with respect to $i$. Inspection of the figure reveals the matrix elements are only significantly correlated up to the second and third neighbours. 

The picture which emerges is of a predominantly tri-diagonal matrix with off-diagonal band element values having a non-negligible probability of being quite small. Viewing this matrix to be precisely a tri-diagonal matrix and with these small off-diagonal values to be identically zero results in a matrix of non-interacting smaller block matrices. This is equivalent to truncating the dislocation-dislocation interaction to a distance of approximately $\lambda_{0}$. In this limit the origin of purely localized eigenmodes becomes manifest, with the eigenmodes consisting of dislocation bunches existing within one $\lambda_{0}$ unit. Restoring the dislocation-dislocation interaction to its full range results in a weak interaction between the block matrices and therefore the possibility of more spatially extended eigenmodes. That both the diagonal and off-diagonal band elements are weakly correlated beyond three element indices, means that at a resolution above that of the strongly interacting block matrices, the full matrix is random. Such a random block tri-diagonal matrix is similar as the one-dimensional random potential problem and the physics of Anderson localization~\cite{Anderson1958}. In one dimension, the localization phenomenon is robust against the strength of the disorder~\cite{Abrahams1979}, and --- barring any long-range correlations between the random matrix elements (not suggested by Fig.~\ref{FigDis1}d and the $1/r^{2}$ dependency of the off-diagonal elements) --- are expected to be localized as seen via the participation number in Fig.~\ref{FigSim}d. Indeed, analysis of the level separation statistics of the eigenvalues reveals strong Poisson behaviour (not shown).

\subsection{Relation between the emerging instability and irreversible plastic strain} \label{SSecPS}
 
When the applied stress equals the critical stress of the plastic event, the unstable eigenmode's eigenvalue becomes identically zero and the linearization procedure breaks down. During the subsequent regime of strong non-linearity, dislocations within the structure evolve according to Eqn.~\ref{EqnEoM}, until a new stable dislocation configuration is found. At this point the plastic event has entirely arrested and the linearization procedure can once again be used. An interesting question is how related is the resulting irreversible plasticity to the unstable mode that triggers it?

\begin{figure}
	\includegraphics[clip,width=0.47\textwidth]{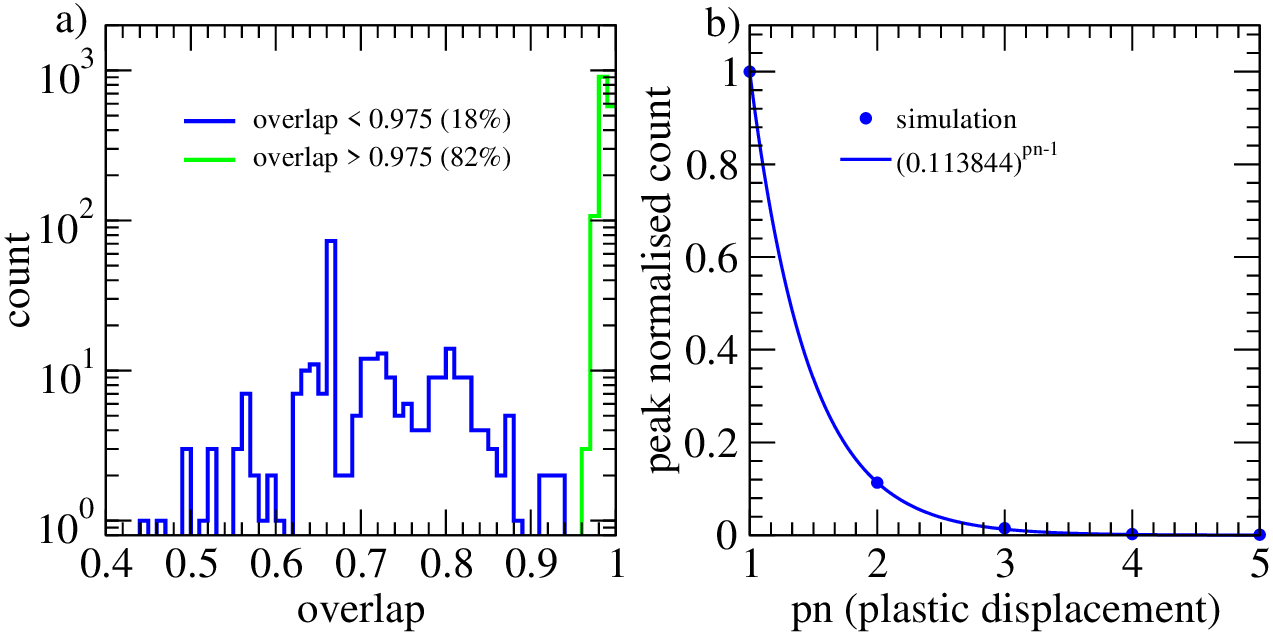}
	\caption{a) Histogram of overlap between the unstable eigenmode and the normalized plastic displacement vector characterizing the resulting irreversible plasticity. b) Peak normalized histogram of the number of dislocations involved in the plastic event obtained from the participation number of the normalized plastic displacement vector (Eqn.~\ref{EqnPN}). Data is derived from approximately two thousand loading simulations for a system size of $d=1028$ $\mu$m. \label{FigPS}}
\end{figure}

To quantify the irreversible plasticity the dislocation displacements (taking into account the periodic boundary conditions) between the positions at the instability and at arrest are found, giving a vector which is then normalized. Taking the dot-product of this normalized plastic displacement vector with the unstable mode gives the overlap and therefore the degree of similarity between the dislocations responsible for the trigger of plasticity and those involved in the actual resulting plasticity. Such an approach has been used to investigate a similar relationship in hard-sphere glasses~\cite{Brito2007} --- see also ref.~\cite{Sharma2014} for the case of magnetic reorientations in an XY spin glass. Fig.~\ref{FigPS}a plots a histogram of this overlap measure derived from several thousand loading simulations performed up to the arrest of the first plastic event for $d=1280$ $\mu$m. The histogram shows two distinct features, where in the first case, there exists a peak (marked in green) close to an overlap of unity, demonstrating that approximately 82\% of the plastic event structures are strongly related to the unstable mode triggering the irreversible plasticity. The remaining 18\% percent of plastic events admit a range of lower overlaps (marked in blue).

Given the eigenmode at the instability involves only one dislocation (Fig.~\ref{FigSim}d), this result indicates 82\% percent of the plastic events are mediated by only the trigger dislocation. This is confirmed through inspection of the participation number (Eqn.~\ref{EqnPN}) of the plastic strain vector. On the other hand, the participation number of those plastic events which have a reduced overlap can have a value ranging between one and five dislocations. Fig.~\ref{FigPS}b plots a histogram of the participation number of the plastic strain vector, in which the bin width is unity and each bin is centred on an integer. The observed trend my be rationalized by considering the probability, $A$, that a dislocation involved in plasticity triggers another dislocation to become mobile, and so on, giving a histogram equal to $A^{n-1}$ where $n$ is the $n$th dislocation involved in the plastic event. Fig.~\ref{FigPS}b includes such a curve for the optimal value of $A=0.113844$, indicating a dislocation involved in plasticity has the probability of $\sim0.1$ to trigger another dislocation. This value is quite compatible with the above observation that approximately 18\% of the plastic events have weak overlap with the unstable eigenmode. The general trends shown here differ little for the considered system sizes down to $d=80$ $\mu$m.

Past work has shown that the present model applied to a dipolar mat geometry exhibits scale-free avalanche behaviour only when the external stress is ``tuned'' to the above yield stress~\cite{Derlet2013}. The transition to macroscopic flow is thus characterized by a mean-field depinning transition~\cite{Fisher1998,Zaiser2006} as discussed by Dahmen and co-workers~\cite{Dahmen2009} --- a result that is also the case for the single slip plane system studied presently. The first irreversible plastic event is therefore expected to not exhibit scale free avalanche behaviour, indeed the statistics of the plastic displacement shown in Fig.~\ref{FigPS}b can not admit power law behaviour in the plastic strain magnitude distribution because $A<1$. This should be compared when approaching yield, and ``tuned'' criticality. Here, $A\rightarrow1$, and the plastic events can involve a very large number of dislocations and therefore a divergent plastic strain.

\section{Discussion} \label{SecDis}

Further insight into the underlying mechanism of the first plastic event can be gained by investigating the dynamical matrix using second order eigenvalue perturbation theory. Treating the off-diagonal elements of the dynamical matrix as a perturbation, Rayleigh-Schr\"{o}dinger gives the lowest eigenvalue as
\begin{equation}
e_{1}\approx\Lambda_{11}+\sum_{i\ne 1}\frac{\left|\Lambda_{1j}\right|^{2}}{\Lambda_{11}-\Lambda_{ii}}, \label{EqnMS}
\end{equation}
where  $\Lambda_{11}$ is the lowest diagonal element. The second term is always negative since $\Lambda_{11}<\Lambda_{ii}$. Two approaches can be taken in assigning the labelling $i$ in the above equation. If the dominant term contributing to the summation arises from the second lowest diagonal element, giving the smallest denominator, then $i$ should reflect the numerical order of the $i$th diagonal element. If, on the other hand, the dominant term originates from the largest off-diagonal element, giving the largest numerator, then $i$ should reflect the numerical order of the spatial distance associated with the $\Lambda_{1i}$ diagonal element. For the present case, it is the latter, so the largest contribution always arises from the term with the largest off-diagonal element and therefore the mode closest to the spatial location of the dislocation associated with the $\Lambda_{11}$ element. This originates from the block structure of the matrix discussed in Sec.~\ref{SSecDM} in which the first plastic event originates from configurations typically involving two or three dislocations within the same $\lambda_{0}$ unit. Because of this, the corresponding $i=2$ diagonal mode $\Lambda_{22}$ is considerably larger than $\Lambda_{11}$.

\begin{figure}
	\includegraphics[clip,width=0.47\textwidth]{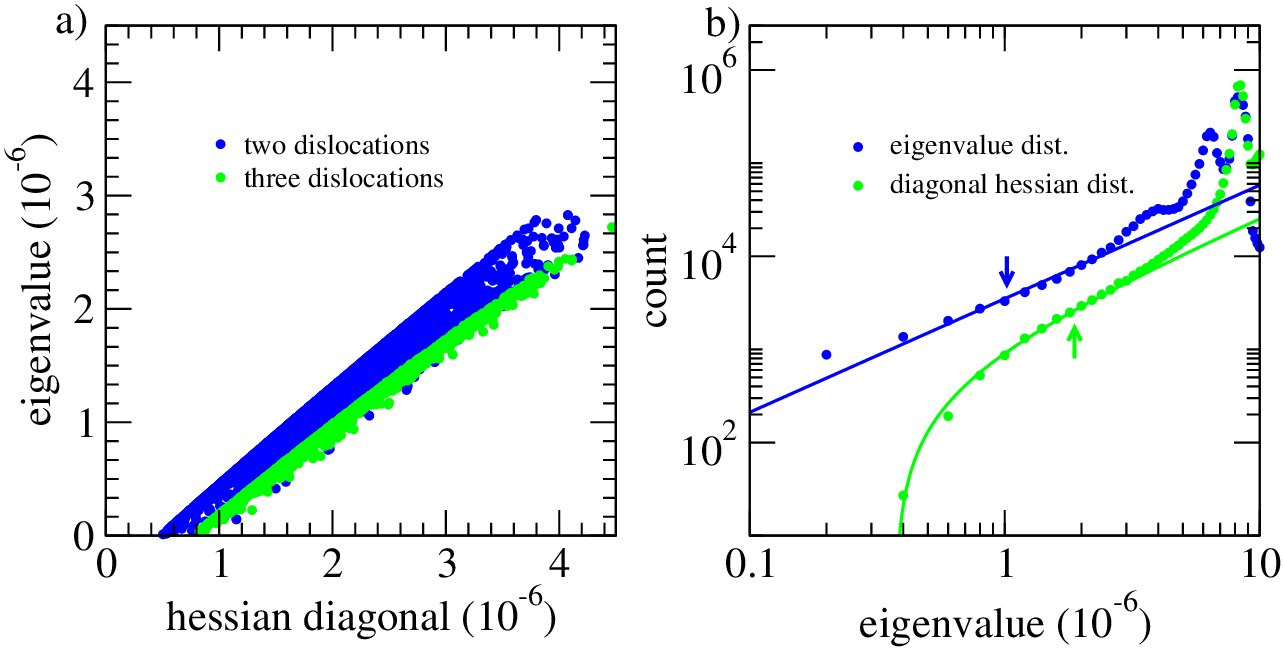}
	\caption{a) Scatter plot of dynamical matrix diagonal versus eigenvalue for the lowest, second lowest and third lowest pair. Those coloured blue/green involve two/three dislocations. b) Low end distributions of the dynamical matrix diagonals and eigenvalues with predicted power law forms. Data is derived from ten thousand initial configurations for a system size of $d=1028$ $\mu$m. \label{FigDis2}}
\end{figure}

Fig.~\ref{FigDis2}a displays a scatter-plot of the lowest eigenvalue versus the lowest dynamical matrix diagonal for $d=1280$ $\mu$m. A strong band of linear correlation is evident. The data is coloured according to the number of dislocations within the $\lambda_{0}$ unit of the dislocation associated with the lowest diagonal element. For the case when the lowest eigenvalue involves two dislocations, lowest order perturbation theory (Eqn.~\ref{EqnMS}) is sufficient to calculate the numerical value of the corresponding eigenvalue. On the other hand, when three dislocations are involved, Eqn.~\ref{EqnMS}, provides only a good estimate and higher-order perturbation theory is needed to obtain an accurate eigenvalue due to the multiple ``scattering'' associated with the two non-negligible off-diagonal elements. Fig.~\ref{FigDis2}a therefore demonstrates two distinct dislocation environments are responsible for the first irreversible plastic event. It is noted that these two mechanisms do not explain the two classes of overlap between the instability mode and the plastic displacement considered in Sec.~\ref{SSecPS}. This suggests the extent of the resulting plasticity has more to do with the surrounding dislocation structure than those dislocations triggering the instability.

The average trend seen in Fig.~\ref{FigDis2}a indicates the distribution of diagonal elements has a hard gap --- $e_{i}\approx\Lambda_{ii}+\Delta\Lambda$ --- which may be estimated by the second term of Eqn.~\ref{EqnMS}. This result is also valid for higher value pairs of diagonal elements and eigenvalues. If, $P(e)\sim e^{\theta}$, then the distribution of diagonal elements must be $P(\Lambda)\sim(\Lambda-\Delta\Lambda)^{\theta}$. What is the value of the exponent $\theta$? The existence of a fold catastrophe gives a direct connection between the first critical stress and the lowest eigenvalues of the dynamical matrix for the configuration prior to loading, $e_{1}\propto\sqrt{\sigma_{\mathrm{c}}}$, giving $\theta=(2-\gamma)/\gamma=1.22$ for the value $\gamma=0.9$ obtained via the average first critical stress versus system size (Fig.~\ref{FigSE}a). The observation that it is not necessarily the lowest eigenvalue which is connected to the critical stress via the fold catastrophe (see fig.~\ref{FigEVal}a) only affects the prefactor of the above scaling, and not the scaling itself. This is because the level spacing of the first few eigenvalues scales with system size in the same way as the lowest eigenvalue. Fig.~\ref{FigDis2}b plots the eigenvalue and diagonal dynamical matrix distributions averaged over 10000 initial configurations for a system size of $d=1280$ $\mu$m along with the corresponding power law distributions (appropriately normalised) using the exponent $\theta=1.22$ and the diagonal element gap (obtained from Fig.~\ref{FigDis2}a). Agreement is seen to be quite good with some deviation close to the hard gap and for the very smallest eigenvalues.

The presence of a gap may be seen without the aid of perturbation theory. In fact, to ensure the dynamical matrix is positive definite (a necessary condition for a stable dislocation configuration), its components must satisfy $|\Lambda_{ij}|\le\sqrt{\Lambda_{ii}\Lambda_{jj}}$ where the diagonal elements are positive non-zero numbers. This is a general result for a positive definite matrix and originates from the property that every principle sub-matrix of the matrix in question must itself be positive definite, including all $2\times2$ sub-matrices. The above inequality then gives
\begin{equation}
\Lambda_{11}\ge\max_{j}\left\{|\Lambda_{1j}|^{2}/\Lambda_{jj}\right\}.\label{EqnMax}
\end{equation}
For the present system, this maximum is always due to the $j$th diagonal element whose dislocation is closest to dislocation `1'. For the case of the smallest possible value of $\max_{j}\left\{|\Lambda_{1j}|^{2}/\Lambda_{jj}\right\}$, the inequality in Eqn.~\ref{EqnMax} limits to an equality, giving the minimum possible value of $\Lambda_{11}$. Indeed, two separate equalities emerge associated with the environments of either two or three dislocations. That this minimum is a finite positive number is due to a unique combination of the repulsive interaction between dislocations (the dislocations cannot be too close to each other), the pinning effect of the static internal field (which tends to bunch dislocations within a single $\lambda_{0}$ unit), and the surrounding dislocation environment. The specific combination of these factors giving the absolute minimum is only approached with a non-negligible probability for large enough system sizes.

The arrows in Fig.~\ref{FigDis2}b indicate the average lowest eigenvalue and the average lowest diagonal element, demonstrating that they are well within the regime for which the power law behaviour is dominant. This is to be expected given the approximately linear scaling between the mean first critical stress versus system size (Fig.~\ref{FigDis2}a). In fact the regime of eigenvalues below that of the hard gap range corresponds (via the fold catastrophe relation) to a critical stress of $\sigma_{\mathrm{c}}\approx0.0008\tau_{0}$, which is well outside the range of Fig.~\ref{FigSE}a. Despite this agreement, it is expected that for large enough system sizes, Weibull~\cite{Weibull1951,Weibull1952,Gumbel1958,Fisher1928,Gnedenko1943} fluctuations around the mean value will be affected in this stress regime: as was noticed for the system size, $d=1280$ $\mu$m, in Ref.~\cite{Derlet2016}.

\section{Concluding Remarks}

The present work has demonstrated the extreme value statistics of the critical stress of the first plastic event in a one-dimensional dislocation dynamics model may be associated with the low-end eigen-structure of the dynamical matrix of the dislocation configuration prior to loading. In particular the critical stress of a given dislocation configuration can be directly related, via a fold catastrophe, to one of the lowest eigenvalues whose corresponding eigenmode will eventually facilitate the instability leading to the first plastic event. Inspection of such eigenmodes and the general statistical properties of the dynamical matrix, reveal them to be localized, involving generally only one to two dislocations. Thus, despite the long range interaction between dislocations, the onset of the very first plastic event for the considered model is a localized structure. It is found that the internal static stress field (with its imposed length and stress scales) plays a dominant role here. 

It is an interesting question, if such a phenomenon may be generalized to two and three spatial dimensions for dislocation networks involving multiple active slip systems and the general phenomenon of screening. In particular, if the characteristic disorder a particular dislocation segment experiences due to all other dislocations is of a type able to induce similar behaviour of the form seen here will be a topic of future work. The work of ref.~\cite{Derlet2016} demonstrates the extreme value statistics approach taken here is applicable to experimental nano-indentation probes of three-dimensional dislocation structure in real materials, suggesting that this might in fact be the case. Furthermore, how this picture might change as a function of the $n$th plastic event might give fundamental insight into the more general nature of micro-plasticity, and the transition to macroscopic yield and plastic flow.

\section{Acknowledgements}

PMD thanks Christopher Mudry and Markus Mueller for helpful discussion.

\end{document}